\def\fracm#1#2{\hbox{\large{${\frac{{#1}}{{#2}}}$}}}

\documentstyle[12pt]{article}

\def\@magscale#1{ scaled \magstep #1}

\catcode`@=11  
\def\un#1{\relax\ifmmode\@@underline#1\else
        $\@@underline{\hbox{#1}}$\relax\fi}
\catcode`@=12

\def\a{\alpha}
\def\b{\beta}

\def\d{\delta}

\def\g{\gamma}

\def\l{\lambda}
\def\m{\mu}

\def\o{\omega}

\def\r{\rho}
\def\s{\sigma}
\def\t{\tau}

\def\z{\zeta}

\def\dslash{\not{\hbox{\kern-2pt $\partial$}}}
\def\Dslash{\not{\hbox{\kern-4pt $D$}}}
\def\pslash{\not{\hbox{\kern-2.3pt $p$}}}
 \newtoks\slashfraction
 \slashfraction={.13}
 \def\slash#1{\setbox0\hbox{$ #1 $}
 \setbox0\hbox to \the\slashfraction\wd0{\hss \box0}/\box0 }
 
\font\ro=cmsy10                          
\def\kcr{{\hbox{\ro \char'170}}}                
\def\ktl{{\hbox{\ro \char'170}}}        
\def\ktr{{\hbox{\ro \char'170}}}        
\def\kbl{{\hbox{\ro \char'170}}}        
\def\kbr{{\hbox{\ro \char'170}}}        

\def\plpl{\raise-2pt\hbox{$\raise3pt\hbox{$_+$}\hskip-6.67pt\raise0.0pt
\hbox{$^+$}\hskip 0.01pt$}}
\def\mimi{\raise-2pt\hbox{$\raise3pt\hbox{$_-$}\hskip-6.67pt\raise0.0pt
\hbox{$^-$}\hskip 0.01pt$}} 

\def\bo{{\raise.15ex\hbox{\large$\Box$}}}               
\def\pa{\partial}                                       
\def\TH{{\raise.2ex\hbox{$\displaystyle \bigodot$}\mskip-4.7mu \llap H\;}}
\def\face{{\raise.2ex\hbox{$\displaystyle \bigodot$}\mskip-2.2mu \llap{$\ddot \smile$}}}                                      

\def\sp#1{{}^{#1}}                              
\def\Bar#1{\overline{#1}}                       
\def\leftrightarrowfill{$\mathsurround=0pt \mathord\leftarrow \mkern-6mu
        \cleaders\hbox{$\mkern-2mu \mathord- \mkern-2mu$}\hfill
        \mkern-6mu \mathord\rightarrow$}
\def\dvec#1{\vbox{\ialign{##\crcr
        \leftrightarrowfill\crcr\noalign{\kern-1pt\nointerlineskip}
        $\hfil\displaystyle{#1}\hfil$\crcr}}}           

\def\fracm#1#2{\hbox{\large{${\frac{{#1}}{{#2}}}$}}}
\def\frac#1#2{{\textstyle{#1\over\vphantom2\smash{\raise.20ex
        \hbox{$\scriptstyle{#2}$}}}}}                   
\def\sfrac#1#2{{\vphantom1\smash{\lower.5ex\hbox{\small$#1$}}\over
        \vphantom1\smash{\raise.4ex\hbox{\small$#2$}}}} 
\def\bfrac#1#2{{\vphantom1\smash{\lower.5ex\hbox{$#1$}}\over
        \vphantom1\smash{\raise.3ex\hbox{$#2$}}}}       
\def\afrac#1#2{{\vphantom1\smash{\lower.5ex\hbox{$#1$}}\over#2}}    

\newskip\humongous \humongous=0pt plus 1000pt minus 1000pt
\def\caja{\mathsurround=0pt}
\def\eqalign#1{\,\vcenter{\openup2\jot \caja
        \ialign{\strut \hfil$\displaystyle{##}$&$
        \displaystyle{{}##}$\hfil\crcr#1\crcr}}\,}
\newif\ifdtup

\def\ref#1{$\sp{#1)}$}

\parskip=\medskipamount                 
\thicklines                         

\def\oldheadpic{                                
        \setlength{\unitlength}{.4mm}
        \thinlines
        \par
        \begin{picture}(349,16)
        \put(325,16){\line(1,0){4}}
        \put(330,16){\line(1,0){4}}
        \put(340,16){\line(1,0){4}}
        \put(335,0){\line(1,0){4}}
        \put(340,0){\line(1,0){4}}
        \put(345,0){\line(1,0){4}}
        \put(329,0){\line(0,1){16}}
        \put(330,0){\line(0,1){16}}
        \put(339,0){\line(0,1){16}}
        \put(340,0){\line(0,1){16}}
        \put(344,0){\line(0,1){16}}
        \put(345,0){\line(0,1){16}}
        \put(329,16){\oval(8,32)[bl]}
        \put(330,16){\oval(8,32)[br]}
        \put(339,0){\oval(8,32)[tl]}
        \put(345,0){\oval(8,32)[tr]}
        \end{picture}
        \par
        \thicklines
        \vskip.2in}
\def\oldtitle#1#2#3#4{\oldheadpic\begin{center}\vglue.5in{\large\bf
#1}\\[.6in]
        {#2}\\[.1in] {\it Department of Physics and Astronomy}\\
        {\it University of Maryland, College Park, MD 20742}\\[.6in]
        Physics Publication \#{#3}\\ {#4}\\[1.5in] {\bf
ABSTRACT}\\[.1in]
        \end{center} \begin{quotation}}                 
\def\oldTitle#1#2#3#4#5#6#7{\oldheadpic\begin{center} \vglue .4in
        {\large\bf #1}\\[.4in]
        {#2}\\[.1in] {\it Department of Physics and Astronomy}\\
        {\it University of Maryland, College Park, MD 20742}\\[.1in]
        {#3}\\[.1in] {\it {#4}}\\ {\it {#5}}\\[.4in]
        Physics Publication \#{#6}\\ {#7}\\[.5in] {\bf ABSTRACT}\\[.1in]
        \end{center} \begin{quotation}}                 
\def\border{                                            
        \setlength{\unitlength}{1mm}
        \newcount\xco
        \newcount\yco
        \xco=-21
        \yco=12
        \begin{picture}(140,0)
        \put(\xco,\yco){$\ktl$}
        \advance\yco by-1
        {\loop
        \put(\xco,\yco){$\kcr$}
        \advance\yco by-2
        \ifnum\yco>-240
        \repeat
        \put(\xco,\yco){$\kbl$}}
        \xco=158
        \yco=12
        \put(\xco,\yco){$\ktr$}
        \advance\yco by-1
        {\loop
        \put(\xco,\yco){$\kcr$}
        \advance\yco by-2
        \ifnum\yco>-240
        \repeat
        \put(\xco,\yco){$\kbr$}}
        \put(-20,13){\tiny University of Maryland Elementary
        ParticlePhysics University of Maryland Elementary Particle
        Physics University of Maryland Elementary Particle Physics}
        \put(-20,-241.5){\tiny The University of Iowa Particle Theory GroupThe University of Iowa Particle Theory Group The University of Iowa  Particle Theory Group The University}
        \end{picture}
        \par\vskip-8mm}
\def\bordero{                                           
        \setlength{\unitlength}{1mm}
        \newcount\xco
        \newcount\yco
        \xco=-31
        \yco=12
        \begin{picture}(140,0)
        \put(\xco,\yco){$\ktl$}
        \advance\yco by-1
        {\loop
        \put(\xco,\yco){$\kclr$}
        \advance\yco by-2
        \ifnum\yco>-240
        \repeat
        \put(\xco,\yco){$\kbl$}}
        \xco=151
        \yco=12
        \put(\xco,\yco){$\ktr$}
        \advance\yco by-1
        {\loop
        \put(\xco,\yco){$\kcr$}
        \advance\yco by-2
        \ifnum\yco>-240
        \repeat
        \put(\xco,\yco){$\kbr$}}
        \put(-20,12){\ooobacdefghidfghghdhededbihdgdfdfhhdheidhdhebaaahjhhdahbahgdedgehgfdiehhgdigicba}
        \put(-20,-241.5){\oooababaighefdbfghgeahgdfgafagihdidihiidhiagfedhadbfdecdcdfagdcbhaddhbgfchbgfdacfediacbabab}
        \end{picture}
        \par\vskip-8mm}
\def\headpic{                                           
        \indent
        \setlength{\unitlength}{.4mm}
        \thinlines
        \par
        \begin{picture}(29,16)
        \put(165,16){\line(1,0){4}}
        \put(170,16){\line(1,0){4}}
        \put(180,16){\line(1,0){4}}
        \put(175,0){\line(1,0){4}}
        \put(180,0){\line(1,0){4}}
        \put(185,0){\line(1,0){4}}
        \put(169,0){\line(0,1){16}}
        \put(170,0){\line(0,1){16}}
        \put(179,0){\line(0,1){16}}
        \put(180,0){\line(0,1){16}}
        \put(184,0){\line(0,1){16}}
        \put(185,0){\line(0,1){16}}
        \put(169,16){\oval(8,32)[bl]}
        \put(170,16){\oval(8,32)[br]}
        \put(179,0){\oval(8,32)[tl]}
        \put(185,0){\oval(8,32)[tr]}
        \end{picture}
        \par\vskip-6.5mm
        \thicklines}
\def\title#1#2#3#4{\border\headpic {\hbox to\hsize{#4 \hfill UMDEPP #3}}\par
        \begin{center} \vglue .5in {\large\bf #1}\\[.6in]
        {#2}\\[.1in] {\it Department of Physics and Astronomy}\\
        {\it University of Maryland, College Park, MD 20742}\\[1.5in]
        {\bf ABSTRACT}\\[.1in] \end{center} \begin{quotation}}  
\def\Title#1#2#3#4#5#6#7{\border\headpic
        {\hbox to\hsize{#7 \hfill UMDEPP #6}}\par
        \begin{center} \vglue .4in {\large\bf #1}\\[.4in]
        {#2}\\[.1in] {\it Department of Physics and Astronomy}\\
        {\it University of Maryland, College Park, MD 20742}\\[.1in]
        {#3}\\[.1in] {\it {#4}}\\ {\it {#5}}\\[.5in] {\bf ABSTRACT}\\[.1in]
        \end{center} \begin{quotation}}                 
\def\endtitle{\end{quotation}\newpage}                  

\def\ad{{\kern0.5pt
                   \alpha \kern-5.05pt
\raise5.8pt\hbox{$\textstyle.$}\kern 0.5pt}}

\def\bd{{\kern0.5pt
                   \beta \kern-5.05pt
\raise5.8pt\hbox{$\textstyle.$}\kern 0.5pt}}

\def\qd{{\kern0.5pt
                   q \kern-5.05pt \raise5.8pt\hbox{$\textstyle.$}\kern0.5pt}}
\def\Dot#1{{\kern0.5pt
                   {#1} \kern-5.05pt
\raise5.8pt\hbox{$\textstyle.$}\kern0.5pt}}

\begin{document}

\def\gfrac#1#2{\frac {\scriptstyle{#1}}
        {\mbox{\raisebox{-.6ex}{$\scriptstyle{#2}$}}}}
\def\gg{{\hbox{\sc g}}}
\border\headpic {\hbox to\hsize{February 2000 \hfill
{UMDEPP 00-049}}}
\par
\par
\setlength{\oddsidemargin}{0.3in}
\setlength{\evensidemargin}{-0.3in}
\begin{center}
\vglue .10in
{\large\bf Superspace Geometrical Realization of the\\
$N$-Extended Super Virasoro Algebra and its Dual\footnote {Supported 
in part by National  Science Foundation Grant PHY-98-02551.}  }
\\[.5in]
C. Curto \footnote{ccurto@fas.harvard.edu}
\\[0.06in]
{\it Department of Physics,
Harvard University\\
Cambridge, MA 02138 USA}\\[.2in]
S. James Gates, Jr.\footnote{gatess@wam.umd.edu}
\\[0.06in]
{\it Department of Physics, 
University of Maryland\\ 
College Park, MD 20742-4111  USA}\\[.1in]
and \\ [.1in] 
V.G.J. Rodgers\footnote{vincent-rodgers@uiowa.edu\,\, }
\\[0.06in]
{\it  Department of Physics and Astronomy, 
University of Iowa\\ 
Iowa City, Iowa~~52242--1479 USA}\\[.6in]

{\bf ABSTRACT}\\[.01in]
\end{center}
\begin{quotation}
{We derive properties of $N$-extended ${\cal {GR}}$ super Virasoro 
algebras.  These include adding  central extensions, identification 
of all primary fields and the action of the adjoint representation 
on its dual.   The final result suggest identification with the 
spectrum of fields in supergravity theories and superstring/M-theory
constructed from NSR $N$-extended supersymmetric ${\cal {GR}}$ 
Virasoro algebras.}

${~~~}$ \newline

${~~~~~~~~~~~~~~~~~~~~}$symmetry, Supergravity
\endtitle

\noindent
{\bf {(I.) Introduction}}  

Recently super-derivations were introduced \cite{gates1}
that extend previous work \cite{GRn2} of 1D, $\aleph_0$-extended  
superspace.  This set of super-derivations is closed under graded 
commutation and contains a super Virasoro-like sub-algebra for 
all values of $N$-extended supersymmetry.  The smallest set of 
the derivations that forms a closed algebra under the action of 
the graded commutator contains the following:
$$
\eqalign{ {~~~~}
G_{\cal A} {}^{\rm I} &\equiv~ i \, \t^{{\cal A} + \fracm 12} \, 
\Big[\, \, \pa^{\rm I} ~-~ i \, 2 \, \zeta^{\rm I} \, \pa_{\t} ~ 
\Big] ~+~ 2 (\, {\cal A} \,+\, \fracm 12 \,) \t^{{\cal A} - 
\fracm 12}  \z^{\rm I} \z^{\rm K} \, \pa_{\rm K} ~~~, \cr
L_{\cal A} &\equiv~ -\, \Big[ \, \t^{{\cal A} + 1} \pa_{\t} ~+~ \fracm
12
({\cal A} \, + \, 1) \, \t^{\cal A} \zeta^{\rm I} \, \pa_{\rm I} ~
\Big]  
~~~, \cr
T_{\cal A}^{\, \rm {I \, J }} &\equiv~ \t^{\cal A} \, \Big[ ~ \z^{\rm I} 
\, \pa^{\rm J} ~-~ \z^{\rm J} \, \pa^{\rm I}  ~\Big] ~~~, \cr
U_{\cal A}^{{\rm I}_1 \cdots {\rm I}_q} &\equiv~ i \, (i)^{ [\fracm q2
]} 
\, \t^{({\cal A} - \fracm {(q - 2)}2 )} \zeta^{{\rm I}_1} \cdots \,
\zeta^{{\rm I}_{q-1}} \, \pa^{{\rm I}_q} ~~~,~~~ q \, = \,  3 , \, \dots
\,
, \, N \, + \, 1 ~~~,  \cr
R_{\cal A}^{{\rm I}_1 \cdots {\rm I}_p} &\equiv~  (i)^{ [\fracm p2 
]} \, \t^{({\cal A} - \fracm {(p - 2)}2 )} \zeta^{{
\rm I}_1} \cdots \, \zeta^{{\rm I}_p} \, \pa_{\t} ~~~~~~,~~~ p \, = \, 2
, 
\, \dots \, , \, N  ~~~,  }  \eqno(1)
$$
for any number $N$ of supersymmetries.  (Our notational conventions
can be found in \cite{gates1}.) These derivations do not depend on a 
specific value of $N$ and can therefore be used for the entire 1D, 
$\aleph_0$ superspace.  For low values of $N$, not all of the generators 
appear. For example, $T_{\cal A}^{\, \rm {I \, J }}$ and $R_{\cal 
A}^{{\rm I}_1 \cdots {\rm I}_p}$ only appear for superspaces with $N$ 
$\ge$ 2. Generically, $U_{\cal A}^{{\rm I}_1 \cdots {\rm I}_q}$ only 
appears for superspaces with $N$ $\ge$ 3.  The indices denoted by 
${\cal A}$, ${\cal B}$, etc. denote the level or mode number of the 
operators.  These types of indices take their values in $Z + \fracm 
12 \oplus Z$.

One of the tasks of this paper is to centrally extend the algebra
generated by the above generators.  We impose the Jacobi identity on
all possible combinations of the generators and find that the centrally
extended algebra is given by\footnote{Some of the results here
contain minor corrections to those given in \cite{gates1}.}
$$
\eqalign{
[~ L_{\cal A} \, , \, L_{\cal B} ~\} \,  &=~ ( \, {\cal A} \, 
- \, {\cal B} \, ) \, L_{{\cal A} + {\cal B}} + \fracm {1}{8} 
\,c \,({\cal A}^3-{\cal A})\, \d_{{\cal A} + {\cal B},0}~~~,\cr
[ ~ L_{\cal A} \,, \, U_{\cal B}^{{\rm I}_1 \cdots {\rm I}_m}
~ \} &=~ - \, [ ~ {\cal B} \, + \, \fracm 12 \, (m-2) \, {\cal 
A} ~] \, U_{\cal A + \cal B}^{{\rm I}_1 \cdots {\rm I}_m} ~~~,
\cr
[ ~ G_{\cal A}{}^{\rm I} \, , \, G_{\cal B}{}^{\rm J} ~\} \, &=~ 
-\, i \, 4 \,  \d^{{\rm {I\, J}}} L_{{\cal A} + {\cal B}}
~-~ i 2 ({\cal A} - {\cal B} ) \, [ \, \, T_{{\cal A} + {\cal B}
}^{\rm {I\, J}} ~+~ 2  ({\cal A} + {\cal B} ) \, U_{{\cal A} + 
{\cal B}}^{\rm {I\, J \,K}}{}_{\rm K} ~ ]   \cr
&{~~~~~} -i c ({\cal A}^2 -\fracm 14 ) \, \d_{{\cal 
A+ B}, 0} \, \d^{\rm I\, J}~~~, \cr
[ ~ L_{\cal A} \, , \, G_{\cal B}{}^{\rm I} ~\} \,  &=~ ( \, 
\fracm 12 {\cal A} \, - \, {\cal B} \, ) \, G_{{\cal A} + {\cal 
B}}{}^{\rm I}~~~, \cr
[ ~ L_{\cal A} \,, \, R_{\cal B}^{{\rm I}_1 \cdots {\rm I}_m} ~ \} 
&=~ -  \,\, [ ~ {\cal B} \, + \, \fracm 12 \, (m - 2) \, {\cal A}
~] \, R_{\cal A + \cal B}^{{\rm I}_1 \cdots {\rm I}_m} \cr
&{~~~~~}-~ \, [ ~  \fracm 
12 \,{\cal A} \, ({\cal A} + 1  ) ~] \,  U_{\cal A + \cal B}^{{\rm 
I}_1 \cdots {\rm I}_m \, J}{}_J ~~~, \cr 
[ L_{\cal A} \, ,\, T_{\cal B}^{\, \rm {I \, J }} \} \, &=~ - \,{\cal B} 
\, \,T_{\cal A+B}^{\, \rm {I \, J }} ~~~, }$$
$$\eqalign{ 
[ ~ R_{\cal A}^{{\rm I}_1 \cdots {\rm I}_m} \,, \, R_{\cal B}^{{\rm 
J}_1 \cdots {\rm J}_n} ~ \} &=~ -\, (i)^{\s(m n)}  \, [ ~ {\cal A} \, -
\, {\cal B} \, - \, \fracm 
12 (m \, - \, n ) ~] \, R_{\cal A + \cal B}^{{\rm I}_1 \cdots {\rm 
I}_m \, {\rm J}_1n \cdots {\rm J}_n} ~~~, \cr
[~ T_{\cal A}^{\, \rm {I \, J }} \,,\,  T_{\cal B}^{\, \rm {K \, L }}
~\} 
&=~ T_{\cal A+B}^{\, \rm {I \, K }} \, \d^{\rm J L}+~ T_{\cal A+B}^{\, 
\rm {J \, L }} \, \d^{\rm I K}-~ T_{\cal A+B}^{\, \rm {I \, L }} \,
\d^{\rm J K}-~ T_{\cal A+B}^{\, \rm {J \, K }} \, \d^{\rm I L} \cr
&{~~~~~}+~{\tilde c} \, ({\cal A}-{\cal B}) (\d^{\rm I K} \d^{\rm J L}\, 
- \d^{\rm I L}  \d^{\rm J K}) \d_{{\cal A} +{\cal B},0} ~~~,\cr
[ ~ G_{\cal A}{}^{\rm I} \,, \, R_{\cal B}^{{\rm J}_1 \cdots {\rm 
J}_m}~ \} &=~  2 \, (i)^{\s(m)} \,[ ~ {\cal 
B} \, + \, (m - 1) \, {\cal A} \, + \, \fracm 12 \, ~]\, R_{\cal A + 
\cal B}^{{\rm I} \, {\rm J}_1 \cdots {\rm J}_m} \cr 
&{~~~~\,}-\, (i)^{\s(m)} \, 
\sum_{r=1}^m \, (-1)^{r -1} \, \d^{I \,  {\rm J}_r} \, R_{\cal A + 
\cal B}^{{\rm J}_1 \cdots {\rm J}_{r-1} \,  {\rm J}_{r+1} \cdots {\rm 
J}_m} \cr  
&{~~~~\,}-\,  (-i)^{\s(m)} \, [~{\cal A} \, +\, \fracm 12 ~] \, U_{\cal
A  + \cal B}^{{
\rm J}_1 \cdots {\rm J}_m \, {\rm I}} \cr
&{~~~~\,}+\, 2  \, (i)^{\s(m)} \, 
[~ {\cal A}^2 \, -\, \fracm 14 ~] \, U_{\cal A  + \cal B}^{I\, {\rm 
J}_1 \cdots {\rm J}_m \, K}{}_K ~~~, \cr
[ ~ G_{\cal A}{}^{\rm I} \, , \, U_{\cal B}^{{\rm J}_1 \cdots {\rm J}_m}
~ \} &=~  2 \, (i)^{\s(m)} \, [ ~ {\cal B} 
\, + \, (m - 2) \, {\cal A} ~] \, U_{\cal A  + \cal B}^{{\rm I} \, {\rm
J}_1 \cdots {\rm J}_m} \cr
&{~~~~\,}- \,2 \, (-i)^{\s(m)} \, [ ~ {\cal A} \, + \, \fracm 12 ~] \, 
\d^{{\rm I} \, {\rm J}_m} \, U_{\cal A  + \cal B}^{ {\rm J}_1 \cdots 
{\rm J}_{m - 1} K} {}_K \cr
&{~~~~\,}-\,(i)^{\s(m)} 
\, \sum_{r=1}^{m - 1} \, (-1)^{r -1} \, \d^{{\rm I} \, {\rm J}_r} \, 
U_{\cal A + \cal B}^{{\rm J}_1 \cdots {\rm J}_{r-1} \, {\rm J}_{r+1} 
\cdots  {\rm J}_m} \cr 
&{~~~~\,} +\, 2 \,(-i)^{\s(m)}\, \d^{{\rm I} \, {\rm J}_m} \, R_{\cal A
+
\cal  B}^{{\rm J}_1 \cdots {\rm J}_{m-1} } ~~~, \cr
[ ~ R_{\cal A}^{{\rm I}_1 \cdots {\rm I}_m} \,, \, U_{\cal B}^{{\rm J}_1 
\cdots {\rm J}_n} ~ \} &=~ (-i)^{\s(m n)} 
\, \sum_{r=1}^m \, (-1)^{r -1} \,\d^{{\rm 
I}_r \, {\rm J}_n} \, R_{\cal A + \cal B}^{{\rm J}_1 \cdots {\rm J}_{n-1
} {\rm I}_1 \cdots {\rm I}_{r-1}  \, {\rm I}_{r+1} \cdots {\rm I}_m}  
\cr
&{~~~~~}+ \, i (i)^{\s(m n)} \, [ ~ {\cal B} \, - \, \fracm 12 (n - 2)
~] \, U_{\cal A  
+ \cal B}^{{\rm I}_1 \cdots {\rm I}_m  \, {\rm J}_1 \cdots {\rm J}_n } 
~~~, \cr
[ ~ U_{\cal A}^{{\rm I}_1 \cdots {\rm I}_m} \,, \, U_{\cal B}^{{\rm J}_1 
\cdots {\rm J}_n} ~ \} &=~-  \,  (i)^{\s(m n)}\, \Big\{ \, \sum_{r =
1}^m \, (-1)^{r-1} 
\, \d^{{\rm I}_m {\rm J}_r} \, \, U_{\cal A + \cal B}^{{\rm I}_1 \cdots 
{\rm I}_{m-1} \, {\rm J}_1 \cdots {\rm J}_{r-1} \, {\rm J}_{r+1} \cdots 
{\rm J}_{n-1} {\rm J}_n }~ \cr
&{~\,}{~~~~} -\, (-1)^{mn} \,  \sum_{r = 1}^m \, 
(-1)^{r-1} \, \d^{{\rm I}_r {\rm J}_n} \, \, U_{\cal A + \cal B}^{{\rm 
J}_1 \cdots {\rm J}_{n-1} \, {\rm I}_1 \cdots {\rm I}_{r-1} \, {\rm 
I}_{r+1} \cdots {\rm I}_{m-1} {\rm I}_m } \, \Big\} ~ ~~ , \cr
[~ T_{\cal A}^{\, \rm {I \, J }}\, , \, G_{\cal B}{}^{\rm K} ~\} \, 
&=~ 2\,(\d^{\rm J K} G_{\cal A+B}{}^{\rm I} \, - \d^{\rm I K} G_{\cal 
A+B}{}^{\rm J}) \cr
&{~~~}~+~ 2 {\cal A} \, (\d^{\rm J K} \, \, U_{{\cal A} + 
{\cal B}}^{\rm {I \, , L}}{}_{\rm L} - \d^{\rm I K} \, \, U_{{\cal A} 
+ {\cal B}}^{\rm {J \, L}}{}_{\rm L}  + \, U_{{\cal A} + {\cal B}}^{
\rm {J \, K \, I\,}} - \, U_{{\cal A} + {\cal B}}^{\rm {I\, K \, J}} 
) ~~~, \cr
[~T_{\cal A}^{\, \rm {I \, J }}\,, \, R_{\cal B}^{{\rm I}_1 \cdots {\rm
I}_p} ~\}  &=~ \sum_{r=1}^p \, (-1)^{r +1} \,( \d^{ {\rm J\, I}_r} \,
R_{\cal A +  B}^{{\rm I} \, {\rm I}_1 \cdots {\rm I}_{r-1} \, 
{\rm I}_{r+1} \cdots {\rm I}_p} - \d^{{\rm I\, I}_r} \,R_{\cal A + 
B}^{{\rm J} \, {\rm I}_1 \cdots {\rm I}_{r-1} \, \, {\rm I}_{r+1} 
\cdots {\rm I}_p})\cr 
&{~~}~~+ i \,(-1)^p {\cal A} \, (  U_{{\cal A+B}
-1}^{\rm {I\, I_1 \cdots I_p \,J }} - U_{{\cal A+B}-1}^{\rm {J \, 
I_1 \cdots I_p \, I }}) ~~~, \cr
[~T_{\cal A}^{\, \rm {I \, J }} \,,\, U_{\cal B}^{{\rm I}_1 \cdots {\rm 
I}_p} ~\}  &=~ \sum_{r=1}^p \, (-1)^{r +1} \,( \d^{ {\rm J\, I}_r} \, 
U_{\cal A +  B}^{{\rm I} \, {\rm I}_1 \cdots {\rm I}_{r-1} \, {\rm 
I}_{r+1} \cdots {\rm I}_p} - \d^{{\rm I\, I}_r} \,U_{\cal A +  B}^{{
\rm J} \, {\rm I}_1 \cdots {\rm I}_{r-1} \, \, {\rm I}_{r+1} \cdots 
{\rm I}_p})\cr
&{~~~~~}+  \, (  \d^{\rm I_p \, I}\, U_{{\cal A+B}-1}^{\rm 
{ I_1 \cdots I_{p-1} \,J }} -\, \d^{\rm I_p \, J}\, U_{{\cal
A+B}-1}^{\rm 
{ I_1 \cdots I_{p-1} \,I }} \, ) ~~~,} \eqno(2)
$$
where the function $\s(m)=0$ if $m$ is even and $-1$ if $m$ is odd.
Here the central extensions $c$ and ${\tilde c}$ are unrelated since
we have only imposed the Jacobi identity.  New constraints will arise
when we restrict to unitary representations.   The algebra exhibits
interesting properties such as a generalization of the SO(N)
generators due to the presence of the ${U}$ and ${ R}$ type
fields.  The nature of these fields will be discussed throughout as
we derive transformation laws.  

One way to understand the operators that arise in this new algebra is 
through methods used to study other infinite dimensional algebras.  In
particular, we will borrow techniques from coadjoint representation to 
help interpret these new generators. The coadjoint representation for
infinite dimensional algebras \cite{loop,kirrilov} has appeared in 
the string theory literature for some time.  Its uses include the 
study of chiral anomalies \cite{rai,alek,weigman}, geometric
quantization 
of the Virasoro group \cite{witten3}, the study of orthogonal field 
theories \cite{emma,lano2,emma2} and recently in relation to ${\rm 
AdS}_3$ quantum gravity \cite{navarro}.  

In this paper we will examine the coadjoint representation of the
superspace geometrical representation (``${\cal {GR}}$'') of the 
extended super Virasoro algebras as well as some other properties.  
Although the algebra is  quite complex, the coadjoint representation of
this particular algebra  can generalize many of the above mentioned
aspects as well as shed  light on the meaning of the new generators and
the spectrum of states that may appear in a supergravity or
superstring/M-theory based  on this algebra.    

\noindent
{\bf {(II.) Primary Fields}}

Before going into the coadjoint representation, we would like to 
identify the primary fields associated with this algebra and their 
associated conformal weights.  Since ${\cal L}$ is the generator 
of diffeomorphisms we can use its action on the other generators 
to determine the tensor properties of the fields.  Let
$$ 
{\cal L}^{\prime} =~ \Big( \, L_{\xi},\,G_{\chi^{\rm I}} {}^{\rm I},
\, T_{t^{\rm {J \, K}}}^{\, \rm {J \, K }} , \, \oplus^{\,}_{\rm 
\{I_q\}} \, U_{{\m}^{\rm \{I_q\}}}^{\rm \{I_q\}} , \,  
\oplus^{\,}_{\rm \{J_q\}} \, R_{r^{\rm \{J_q\}}}^{\rm \{J_q\}} ; 
\a \, \Big) ~~~, \eqno(3) $$
represent the generators with generic functions and $\oplus_{\rm 
\{I_q\}}$ represents the direct sum over all distinct generators.
Then from the algebra we see that 
$$\eqalign{
[~ (L_\xi, \a) \,,\, (  L_{\zeta}, {\beta}) ~ \} &=~ ( L_{\xi' \zeta 
-\xi \zeta'  },\, \fracm c{i 2 \pi}\int(\xi'' \zeta' - \zeta'' \xi') 
d\,x) ~~~, \cr 
[~ L_\xi \,, \, G^{\rm I}_{\chi^{\rm I}} ~\} &=~ G^{\rm I}_{(-\xi
(\chi^{\rm I})' +\fracm 12 \xi' \chi^{\rm I})} ~~~, \cr
[ ~ L_\xi \,,\,  T^{\rm{ R} { S}}_{t^{\rm{ R}{ S}}} ~\} &=~ T^{\rm{R} 
{S}}_{ (-\xi\,({t}^{\rm{ R}{S}})')}  ~~~, }$$
$$ \eqalign{
[~ L_\xi \, , \, U^{\rm \{V_r\}}_{w^{\rm \{V_r\}}} ~\} &=~ U^{\rm 
\{V_r\}}_{ (-\xi (w^{\rm \{V_r\}})'  -\fracm 12 (r-2) \xi' w^{\rm 
\{V_r\}})}  ~~~, \cr
[~  L_\xi \,,\,  R^{\rm \{ T_r \}}_{\rho^{\rm \{T_r\}}} ~ \} &=~  
R^{\rm \{T_r\}}_{ (-({\rho}^{\rm \{T_r\}})'\xi - \fracm 12 (r-2) 
\xi' ({\rho}^{\rm \{T_r\}}))  }-\fracm i2  \, U^{\rm \{ T_r
\}}_{(\xi''{\rho}^{\rm \{T_r\}})} ~~~, } \eqno(4)
$$
determines the transformation laws of the functions. 
In the above, we have 
suppressed the Grassman indices.  For example 
$ {w^{\rm {\bar V_1} \cdots {\bar V_n}}}$ the function  associated with
the ${U}$ generators may be written  as $w^{\{\rm V_m\}}$ or simply as
$w$.  From the coefficient of the $\xi'$ summand in the transformation
laws we can write down the conformal weight which is also the rank of
the  one dimensional tensors.  The quantity $\xi$ is a rank one
contravariant tensor field, $\chi^{\rm I}$ is a spin half field and
$t^{\rm R S}$ is a scalar field.   This is to be expected from these
fields.  However, notice that $w^{\rm \{V_r\}}$ transforms with
conformal weight $\fracm 12 (r-2)$ where $r$ takes values from $3$ to 
$N$ when there are $N$  supersymmetries which corresponds to a tower 
of $N-2$ fields.  The quantity $ {\rho}^{\rm \{T_r\}}$ to appears 
to transform as a rank $\fracm 12 (r-2)$ tensor modulo the inhomogeneous
term.  However, it is this inhomogeneous term that keeps these fields
from transforming  like tensors.  Since the transformations of $w^{\rm
\{V_r\}}$ and $ {\rho}^{\rm \{T_r\}}$ are entangled, a natural question
to ask is what linear combination of generators produces tensors or in
the language of conformal field theory which generators are primary.  

To answer this let us consider the generators
$$ {\cal Q}^{\rm I_1 \cdots I_p }_{\cal A} \, =\, \t^{\cal A}
\z^{\rm I_1} \cdots \z^{\rm I_p} \partial_\t ~~~,~~~
{\cal P}^{\rm I_1 \cdots  I_{p+1}}_{\cal A} \, = \, \t^{\cal A}
\z^{\rm I_1} \cdots \z^{\rm I_p} \partial ^{\rm I_{p+1}} 
~~~. \eqno(5) $$
These form a closed algebra among themselves and are used to
facilitate the computations below.  We can write the previous 
generators as
$$\eqalign{
&{ L}_{\cal A} = -{\cal Q}_{{\cal A}+1} - \fracm 12 ({\cal A} 
+ 1) {\cal P}^{\rm I}_{\cal A}{}_{\rm I} ~~~, \cr
&{ G}^{\rm I}_{\cal A} = i {\cal P}^{\rm I}_{{\cal A} + \fracm 
12} + 2 {\cal Q}^{\rm I}_{{\cal A} + \fracm 12} + 2 ({{\cal A} + 
\fracm 12}) {\cal P}^{\rm I\, K}_{{\cal A} - \fracm 12 }{}_{\rm K}
~~~, \cr
&{ T}^{\rm I\, J}_{\cal A}  = {\cal P}^{I\, J}_{\cal A} - {\cal
P}^{J \, I}_{\cal A} ~~~, \cr
&{ R}^{\rm I_1 \cdots I_p}_{\cal A} = \, i^{[\fracm {\rm p}2]} 
{\cal Q}^{\rm I_1 \cdots I_p}_{{\cal A} - \fracm {\rm p}2 +1}, ~~~
({\rm p} = 2 , \cdots, N) ~~~, \cr
&{ U}^{\rm I_1 \cdots I_q}_{\cal A} = i \, (i)^{[\fracm {\rm q}2 ]} 
\, {\cal P}^{\rm I_1 \cdots I_q}_{{\cal A} - \fracm {\rm p}2 + 1}, 
~~~ ({\rm q} = 3 , \cdots, N+1) ~~~.} \eqno(6)
$$
Let ${\cal F}^{\rm I_1 \cdots I_p}_{\cal A}$ be a primary generator. 
Then by definition for some particular mode dependent $\l$ this
generator satisfies
$$[ L_{\cal A}, \, {\cal F}^{\rm I_1 \cdots I_p}_{\cal B} 
\,] \,=\, - \l({\cal A,B},{\rm p}) {\cal F}^{\rm I_1 \cdots 
I_p}_{\cal A+B} ~~~, \eqno(7)
$$
for fixed number of indices p.  For each value of p 
(assuming that p is greater than 2), ${\cal F}^{\rm I_1 \cdots 
I_p}_{\cal A}$ can be generically written as 
$${\cal F}^{\rm I_1 \cdots I_p}_{\cal A} = c_0({\cal A}+1) {\cal 
Q}^{\rm I_1 \cdots I_p }_{\rm A+1} ~+~ c_1({\cal A}+1) {\cal P}^{\rm I_1 
\cdots I_{p}}_{{\cal A}+1} ~+~ c_2({\cal A}) {\cal  P}^{\rm I_1 
\cdots  I_{p} \, L}_{\cal A}{}_{\rm L}  ~~~, \eqno(8) $$ 
which gives us three possible mode dependent coefficients to compute, 
viz $\, c_0, c_1, $ and $c_2$.  From the commutation relations of
${\cal Q}^{\rm I_1 \cdots I_p}_{\cal A}$ and ${\cal P}^{\rm I_1 \cdots
I_{p}}_{\cal A}$ the conditions for a primary generator are
$$\eqalign{
\l c_0({\cal A+B}+1) &= c_0({\cal B} + 1) [({\cal B-A} )+ \fracm 12 
{\rm p}({\cal A}+1)] ~~~, \cr
\l c_1({\cal A+B}+1) &= c_1({\cal B} + 1) [({\cal B-A} )+ \fracm 12 
{\rm p}({\cal A}+1)] ~~~, \cr
\l c_2({\cal A+B})&=-c_0({\cal B}+1)[ \fracm {\rm p}2 ({\cal
A}+1)] ~~~.
}   \eqno(9) $$
There are three classes of solutions. \newline ${~~}$ 
\noindent
${~~~}$ 1. Class 1: \newline \indent
${~~\,}$ Setting $c_0 =1, \, c_1=0$ and $c_2({\cal A}) = a_1 {\cal A} 
+ a_2$ we find that
$$ 
{\cal F^{\rm 1}}^{\rm I_1 \cdots I_p}_{\cal B}\,=\,{\cal Q}^{\rm I_1
\cdots I_p }_{{\cal B}+1} + (\fracm {{\cal B}+1}{2-{\rm p}})\,{\cal
P}^{\rm I_1 \cdots  I_{\rm p}\, L}_{\cal B}{}_{\rm L} ~~~,
\eqno(10) $$
${~~~~~~~\,}$ is a primary field with $\l = ({\cal B} +\fracm {\rm p}2)
- {\cal A}(1 - \fracm {\rm p}2).$  This can be rewritten ${~~~~~~~\,}$ 
${~~~~~~~\,}$ in terms of the original generators as
$$
{R}^{\rm I_1 \cdots I_p}_{{\cal B}+ \fracm {\rm p}2} - i (\fracm
{{\cal B}+1}{2-{\rm p}})\, {U}^{\rm I_1 \cdots I_p\, L}_{{\cal
B}+\fracm p2 -1}{}_{\rm L}~~~, ~~~~~({\rm p}\ne 2) \eqno(11)
$$ 
${~~~~~~~\,}$ 
is a primary field. \newline ${~~}$ 
\noindent
${~~~}$ 2. Class 2: \newline \indent
${~~\,}$ Setting $c_0 =0, \, c_1=1$ and $c_2({\cal A}) = a_1 {\cal A} 
+ a_2$ we find that
$$
{\cal F^{\rm 2}}^{\rm I\,J}_{\cal B}\,={\cal P}^{\rm I J }_{{\cal
B}+1} + ( {\cal B}+1)\,{\cal P}^{\rm I\,J\,K}_{\cal B}{}_{\rm K}
~~~.
\eqno(12) $$
${~~~~~~~\,}$ 
In this case ${\rm p}=2$ was forced as a condition, thus the 
above $\l$ simplifies ${~~~~~~~\,}$ ${~~~~~~~\,}$ to $\l = 
{\cal B}+1$. \newline ${~~}$ 
\noindent
${~~~}$ 3. Class 3: \newline \indent
${~~\,}$ Setting $c_2 =0, \, c_1=1$ and $c_0({\cal A}) = a_1 {\cal 
A} + a_2$ we find that
$$
{\cal F^{\rm 3}}^{\rm I_1 \cdots I_p}_{\cal B}\,={\cal P}^{\rm I_1
\cdots I_p }_{{\cal B}+1} ~~~, \eqno(13)$$
${~~~~~~~\,}$ 
is a primary field.  In this case $c_0=0$ was forced as a condition.

From these three solutions we can find all the primary fields of the
original algebra.
$$\eqalign{
{L}_{\cal B} &=  -{\cal F^{\rm 1}}_{\cal B}~~~,~~~~~~
~~~~~~~~~~~~\,({\rm p}=0) \cr
{G}^{\rm I}_{\cal B} &= 2{\cal F^{\rm 1}}^{\rm I}_{{\cal B}-\fracm
12}\, + i {\cal F^{\rm 3}}^{\rm I}_{{\cal B}-\fracm 12}~~~,~~({\rm
p}=1) \cr
{T}^{\rm I\, J}_{\cal B}  &= {\cal F^{\rm 3}}^{\rm I\, J}_{{\cal
B}-1} - {\cal F^{\rm 3}}^{\rm J\, I}_{{\cal B}-1}~~~, ~~~~~
~\, ({\rm p}=2) }$$
$$\eqalign{
{U}^{\rm I_1 \cdots I_p}_{{\cal B}+\fracm {\rm p}{2}} &=
\, i i^{[\fracm {\rm p}2]} {\cal F^{\rm 3}}^{\rm I_1 \cdots I_p}_{\cal
B}~~~,~~~~~~~~~~~\, ({\rm p} \ge 3) \cr
{R}^{\rm I_1 \cdots I_p}_{{\cal B}+ \fracm {\rm p}2} & -  i (\fracm
{{\cal B}+1}{2-{\rm p}})\, {U}^{\rm I_1 \cdots I_p\, L}_{{\cal
B}+\fracm p2 -1}{}_{\rm L} =
(i)^{[\fracm {\rm p}2 ]} \,{\cal F^{\rm 1}}^{\rm I_1 \cdots I_p}_{{\cal
B}}~~~,~~~~~~~~~~ ({\rm p} \ge 3) ~~~.
} \eqno(14)
$$
We note that ${\cal R}^{I\,J}_{\cal A}$ for no value of ${\cal A}$
admits a primary field.  Stated in a slightly different way, the set
of generators given by 
$$
\eqalign{ {~~~~}
G_{\cal A} {}^{\rm I} &\equiv~ i \, \t^{{\cal A} + \fracm 12}  \,
\Big[\, 
\, \pa^{\rm I} ~-~ i \, 2 \, \zeta^{\rm I} \, \pa_{\t} ~ \Big] ~+~ 2 (\, 
{\cal A} \,+\, \fracm 12 \,) \t^{{\cal A} - \fracm 12}  \z^{\rm I}
\z^{\rm 
K} \, \pa_{\rm K} ~~~, \cr
L_{\cal A} &\equiv~ -\, \Big[ \, \t^{{\cal A} + 1} \pa_{\t} ~+~ \fracm
12
({\cal A} \, + \, 1) \, \t^{\cal A} \zeta^{\rm I} \, \pa_{\rm I} ~
\Big]  
~~~, \cr
T_{\cal A}^{\, \rm {I \, J }} &\equiv~ \t^{\cal A} \, \Big[ ~ \z^{\rm I} 
\, \pa^{\rm J} ~-~ \z^{\rm J} \, \pa^{\rm I}  ~\Big] ~~~,  \cr
U_{\cal A}^{{\rm I}_1 \cdots {\rm I}_q} &\equiv~ i \, (i)^{ [\fracm q2
]} 
\, \t^{({\cal A} - \fracm {(q - 2)}2 )} \zeta^{{\rm 
I}_1} \cdots \, \zeta^{{\rm I}_{q-1}} \, \pa^{{\rm I}_q} ~~~,~~~ q \, =
\, 
3 , \, \dots \, , \, N \, + \, 1 ~~~,  \cr
{\cal R}{}_{\cal A}^{{\rm I}_1 \cdots {\rm I}_p} &\equiv~  (i)^{ [\fracm
p2 ]} \, \t^{({\cal A} - \fracm {p}2 )} \zeta^{{\rm I}_1} \cdots \, 
\zeta^{{\rm I}_p} \, [ \, \t \pa_{\t}  ~+~ ({{\cal A} + 1 \over {p - 
2}}) \z^L \pa_L ~] ~~~~,~~~ p \, = \, 3 , \, \dots \, , \, N ~~~, \cr
R_{\cal A}^{{\rm I} {\rm J}} &\equiv~ i \, \t^{{\cal A} 
} \zeta^{{ \rm I}} \zeta^{{\rm J}} \, \pa_{\t} ~~~~~~,
}  \eqno(15)
$$
possesses only one non-primary generator, namely $R_{\cal A}^{{\rm I}
{\rm J}}$.  We will refer to this basis as the ``almost primary basis'' 
for the ${\cal {GR}}$ super-Virasoro algebra.

\noindent
{\bf {(III.) The Coadjoint Representation}}

In this paper we will examine the coadjoint representation of the
superspace geometrical representation of the extended super Virasoro
algebras as well as some other properties.   Although the algebra is 
quite complex the coadjoint
representation of this particular algebra can generalize many of the
above mentioned aspects as well as shed light on the meaning of the
new generators in the algebra.  

\noindent
{\bf {(III.a)An Example: }}

To begin we will use the semi-direct product of the Virasoro 
algebra and an affine Lie algebra on the circle to fix the notation and
familiarity of the coadjoint representation.   In this case we have 
an affine Lie algebra associated with the loop group G together with 
the Virasoro algebra given by
$$
\eqalign{ 
 [ J^{\alpha}_N, J^{\beta}_M] & = i f^{\alpha\beta\gamma} J^\gamma_{N+M}
+ 
N k \delta_{M+N,0} \delta^{\alpha\beta} ~~~, \cr 
[ L_N,J^{\alpha}_M] & = - M J^{\alpha}_{M+N} ~~~, \cr
[ L_N,L_M ] &= (N-M) L_{N+M} + {{\hat c}\over 12} (N^3-N) \delta_{N+M,0}
~~~, } 
\eqno(16)$$
where ${\hat c} = {2 k {\rm Dim}(G)\over 2k + c_v}$, ${\rm Dim}(G)$ is
the 
dimension of the group and $c_v$ is the value of the quadratic Casimir
in the adjoint representation.  Let $\bigl( L_A, J^\beta_B, \rho\bigr)$ 
denote a centrally extended adjoint vector.  Then from the commutation 
relations above one may write the adjoint action on the adjoint vectors 
as
$$\eqalign{
\bigl( L_A, J^\beta_B, \rho\bigr) \ast \bigl( L_{N'}, J^{\alpha '}_{M'},
\mu\bigr) = &\cr
\bigl( (A-N') L_{A+N'},
-M' J^{\alpha '}_{A+M'} +B J^\beta_{B+N'} +i f^{\beta \alpha ' \lambda} 
J^\lambda_{B+M'}& ,\cr
 {{\hat c}\over 12} (A^3-A) \delta_{A+N',0} &+ B k \delta^{\alpha
'\beta} 
\delta_{B+M',0}\bigr)~~~.} \eqno(17)
$$ 

Now let $\bigl( {\tilde L_N}, {\tilde J^\alpha_M}, {\tilde \mu} \bigr)$
denote an element of the dual space of the algebra and let  
$$
\bigl\langle\bigl( {\tilde L_N}, {\tilde J^\alpha_M}, {\tilde \mu}
\bigr) 
\bigr\vert  \bigl( L_{N'}, J^{\alpha '}_{M'}, \mu ' \bigr)\bigr\rangle 
~=~ \delta^{N,N'} + \delta^{\alpha,\alpha '} \delta_{M, M'} + \mu
{\tilde 
\mu} ~~~, 
\eqno(18) $$
define a suitable pairing.  
By requiring that this pairing be an invariant under the action of any
of the adjoint elements, say $\bigl( L_A, J^\beta_B, \rho\bigr)$,
the coadjoint representation can be defined.   The adjoint action acts 
as a derivation  so that by Leibnitz rule one has
$$
\eqalign{  
&\bigl\langle \bigl( {\tilde L_N}, {\tilde J^\alpha_M}, {\tilde \mu}
\bigr) 
\bigr\vert
\bigl( L_A, J^\beta_B, \rho\bigr) \ast \bigl( L_{N'}, J^{\alpha '}_{M'},
\mu\bigr) \bigr\rangle =\cr
 - &\bigl\langle \bigl( L_A, J^\beta_B, \rho\bigr) \ast \bigl( {\tilde
L_N}, 
{\tilde J^\alpha_M}, {\tilde \mu} \bigr) \bigr\vert 
\bigl( L_{N'}, J^{\alpha '}_{M'},\mu\bigr) \bigr\rangle ~~~.} \eqno(19)
$$    
Thus the transformation properties of the coadjoint vectors are
defined through, 
$$
\eqalign{ 
& \bigl( L_A, J^\beta_B, \rho\bigr) \ast \bigl( {\tilde L_N}, {\tilde 
J^\alpha_M}, {\tilde \mu}\bigr)  = \cr
& \bigl( -(2A-N) {\tilde L_{N-A}} -B \delta^{\alpha\beta} {\tilde
L_{M-B}}
-{{ \tilde \mu} {\hat c}\over 12} (A^3-A) {\tilde L_{-A}},\cr
& (M-A) {\tilde J^\alpha_{M-A}} -i f^{\beta\nu\alpha} {\tilde J^{\nu}_{
M-B}} - {\tilde \mu} B k {\tilde J^\beta_{-B}}, 0 \bigr) ~~~.} 
\eqno{(20)}$$
Instead of using components, let us write $F = (f(\theta),{\hat
h}(\theta),a)$
as an arbitrary adjoint vector and $B=(b(\theta),h(\theta),\mu)$ as an 
arbitrary coadjoint vector, where $f, {\hat h}, b,$ and $h $ are
functions.
For the algebra we choose the realization, 
$$ \eqalign{ L_N& = i \exp{(i N\theta)} \partial_\theta ~~~, \cr
J^\alpha_N& = \tau^\alpha \exp{(i N \theta)} ~~~,} \eqno(21)
$$
and normalize the generators so that ${\rm Tr}( \tau^\alpha \tau^\beta)=
\delta^{\alpha\beta}$. 
Then Eq.(20) may
be written as
$$\eqalign{
&\delta_F B \equiv (f(\theta),{\hat h}(\theta),a) \ast (b(\theta),
h(\theta), \mu) = \cr
-&\bigl( 2 f'b+b'f+i  {{\hat c} \mu\over 12} f''' + {\rm Tr}[ h {\hat
h}'], 
 h'f+hf' + [{\hat h} h - h {\hat h}] + i k \mu {\hat h}', 0 \bigr)
~~~,
}\eqno(22)
$$
where $'$ denotes $\partial_\theta$. The above equation provides an
interpretation of the adjoint elements and coadjoint elements in terms 
of physical fields in one dimension \cite{emma,emma2}.  We already know 
that the Virasoro sector transforms functions as one dimensional
\newpage \noindent coordinate transformations (up to central
extension).  
For example 
$b$ transforms as a rank two tensor field in one dimension where the
infinitesimal coordinate transformation is given by $f$.  From the
second
element of the triplet in Eq.(22), on sees that the  function $h$
transforms
as a one  dimensional gauge field with gauge parameter ${\hat h}$.  The
$
h'f+hf'$ contribution to the transformation of $h$ simply shows that
the field $h$ transforms as a rank one covariant tensor.  The peculiar
transformation is the $Tr[ h {\hat h}']$ that appears in the
transformation of $b$.  This suggests that this rank two tensor can be
shifted by fields built purely from the gauge sector.  In
\cite{emma,emma2} such terms are interpreted as coming from an
interaction Lagrangian.  In any case the relationship between
different members of the algebra juxtaposed to the dual space becomes
manifest through the coadjoint representation.  

Those adjoint vectors, $F$, that leave $B$ invariant will generate the 
{\it isotropy group} for $B$.  Setting Eq.(1a) to zero determines the 
isotropy equation for $B$.  Equation (1a) then determines the tangent
space on the orbit of $B$.  Thus for coadjoint elements $B_1$ and $B_2$, 
we may construct the symplectic
two form by writing 
$$
\Omega_B(B_1,B_2) = \langle B \mid [ F_1,F_2] \rangle ~~~, \eqno(23)
$$ 
where for example $\delta_{F_1} B = B_1$.  In \cite{emma,emma2} the
equation of isotropy is related to constraint equations that come from
a two dimensional field theory.

\noindent
{\bf {(III.b) $N$-Extended ${\cal {GR}}$ Super Virasoro Algebra Dual
Space}}

${~~~~}$ To proceed we will let ${\Bar {\cal L}}$ denote a generic 
coadjoint vector and let ${{\cal L}}$ and ${\cal L}^{\prime}$ denote 
adjoint vectors. The pairing $<  {\Bar {\cal L}}| {{\cal L}} >$ is 
an invariant so we require ${\cal L}^{\prime} <  {\Bar {\cal L}}| 
{{\cal L}} > = 0$.  By Leibnitz rule this means that $< {\cal 
L}^{\prime} * {\Bar {\cal L}}| {{\cal L}} > + <  {\Bar {\cal L}}| 
{\cal L}^{\prime} * {{\cal L}} > = 0$.  It is from here that we 
can extract how $ {\cal L}^{\prime} * {\Bar {\cal L}}$ acts. The 
quantity ${\cal L}^{\prime} * {\Bar {\cal L}}$ will carry the dual 
space back into itself.  In our notation we will use as a basis 
for the N-extended Super Virasoro Algebra
$$  \eqalign{
{\cal L}^{\prime} &=~ \Big( \, L_{a},\,G_{b} {}^{\rm I},\, T_{c}^{\, 
\rm {J \, K }} , \, \oplus_{\rm \{ I_q\}} \, U_{\{d_q\}}^{\rm\{ I_q\}}
\oplus_{\{I_p\}} \, R_{\{e_q\}}^{\{{\rm J}_q\}} ; \a \, \Big) ~~~, \cr  
{\cal L} &=~ \Big( \, L_{z}, \,G_{\g}{}^{\rm Q},\, T_{X}^{\, \rm {R 
\, S }} , \, \oplus_{ \{{\rm V}_l\}} \, U_{\{w_l\}}^{\{{\rm V}_l\}}
\oplus_{\{T_m\}} \, R_{\{h_m\}}^{\{{\rm T}_m\}}  \, ; \b \Big) ~~~, \cr  
{\Bar {\cal L}} &=~ \Big( \, {\Bar L}{}_{\bar z}, \,{\Bar G}{}_{{\bar
\g}}
{}^{\bar {\rm Q}},\, {\Bar T}{}_{\bar X}^{\, \rm {{\bar R} \, {\bar S}
}}
, \, \oplus_{\{{\rm \bar V}_{\bar \ell}\}} \, {\bar U}{}_{\{{\bar
W}{}_{\bar
\ell}\}
}^{\{{\bar {\rm V}}{}_{\bar \ell}\}} , \, \oplus_{\{{\bar T_m}\}} \,
{\Bar R}{}_{{\bar h}_{\bar m}}^{{\{\bar {\rm T}{}_m}\}}  \, ; {\bar \b}
\Big)  ~~~, 
} \eqno(24)$$

We will write a generic functions $f = f_a {\t}^a$ and  $z = z_b
{\t}^{b + 1/2}$ and realize the Virasoro generators as 
$$ \eqalign{
L_{\xi} ~\equiv~ \xi^{A + 1} \, L_{A} &=~ - \Big[ ~(\xi^{A + 1} {\t}^{A
+1} )
\pa_{\t} ~+~ \fracm 12 \xi^{A + 1} {\t}^{A} \zeta^{\rm I}
\pa_{\rm I} ~ \Big] \cr
&=~ - \Big[ ~ \xi \pa_{\t} ~+~ \fracm 12 \xi {\t}^{A} \zeta^{\rm I}
\pa_{\rm I} ~ \Big] ~~~.
} \eqno(25)$$
We will use the subscript of the generators and dual for a
generic function of $\t$.  Each generator and dual element will have a 
specific function and how these functions transform under the action
of specific generators is the aim of this paper.  

Since the action will come from ${\cal L}^{\prime} * {\Bar {\cal L}}$ we
will denote
a generic function as
$$  \eqalign{
{\cal L}^{\prime} &=~ \Big( \, L_{\xi},\,G_{\chi^{\rm I}} {}^{\rm I},\, 
T_{t^{\rm {J \, K}}}^{\, \rm {J \, K }} , \, \oplus \, 
U_{{\m}^{{\rm I}_1 \cdots {\rm I}_q}}^{{\rm I}_1 \cdots {\rm I}_q} , \,  
\oplus \, R_{r^{{\rm J}_1 \cdots {\rm J}_q}}^{{\rm J }_1 
\cdots {\rm J}_q} ; \a \, \Big) ~~~, \cr   
{\Bar {\cal L}}  &=~ \Big( \, {\Bar L}{}_{D},\, {\bar G}_{\psi^{{\bar 
{\rm Q}}}}^{\bar {\rm Q}},\,  {\Bar T}{}_{ {\t}^{{\rm R} \,{\rm S}}}^{\, 
{\bar {\rm R}} {\bar {\rm S}} } , \, \oplus \, {\Bar 
U}{}_{{\o}^{{\bar {\rm V}}_1 \cdots {\bar {\rm V}}_q}}^{{\bar {\rm V}}_1 
\cdots {\bar {\rm V}}_q}, \, \oplus \, {\Bar
R}{}_{{\r}^{{\bar 
{\rm J}}_1 \cdots {\bar  {\rm J}}_p}}^{{\bar {\rm J }}_1 \cdots {\bar 
{\rm J}}_p} ; {\bar \b} \, \Big)   ~~~.  } \eqno(26)$$

The coadjoint action is quite tedious but we can organize the
computation by examining the outcome of each of the commutation
relations in the adjoint representation.  Below is a table that
symbolically will summarize our results. In the notation below
$ ~~L \, * \, L ~~$ is just the commutator of two arbitrary Virasoro 
generators while $ ~~ L \, * \, {\Bar L} ~\to~ {\Bar L} $ is the
coadjoint action from an application of the Virasoro generator on its 
dual ${\Bar L}$ that maps back into the duals of the Virasoro
generators.  Multiple entries in the second column correspond to the
different coadjoint actions that can be extracted from the commutator
in the first column.  

\renewcommand\arraystretch{1.00}
\begin{center}
\begin{tabular}{|c|c|}\hline
\multicolumn{2}{|c|}{ \bf Table 1} \\ \hline\hline
${\rm {Commutator}}$ & ${\rm {Co-adjoint}}~{\rm {Action(s)}}$ 
\\ \hline 
$ ~~L \, * \, L ~~$ &  $ ~~ L \, * \, {\Bar L} ~\to~ {\Bar L}
~~$  \\ \hline
$ ~~L \, * \, G ~~$ &  $ ~~ L \, * \, {\Bar G} ~\to~ {\Bar G}
~~$  \\ \hline
$ ~~L \, * \, T ~~$ &  $ ~~ L \, * \, {\Bar T} ~\to~ {\Bar T}
~~$  \\ \hline
$ ~~L \, * \, U ~~$ &  $ ~~ L \, * \, {\Bar U} ~\to~ {\Bar U}
~~$\\ \hline
$ ~~L \, * \, R ~~$ &  $ ~~ L \, * \, {\Bar R} ~\to~ {\Bar R}
~~,~~ L \, * \, {\Bar U} ~\to~ {\Bar R}$  \\ \hline\hline
$ ~~G \, * \, L ~~$ &  $G \, * \, {\Bar G} ~\to~ {\Bar L} 
~~$  \\ \hline
$ ~~G \, * \, G ~~$ &    ~$ ~~ G \, * \, {\Bar L} ~\to~ {\Bar G}
~~,~~ G \, * \, {\Bar T} ~\to~ {\Bar G}, ~~ G \, * \, {\Bar U} ~\to~ 
{\Bar G}$ \\ \hline
$ ~~G \, * \, T ~~$ &  $ ~~ G \, * \, {\Bar G} ~\to~ {\Bar T},~ G \, 
* \, {\Bar U} ~\to~ {\Bar T},~~
~~$  \\ \hline
$ ~~G \, * \, U ~~$ &  $ ~~ G \, * \, {\Bar U} ~\to~ {\Bar U}
,~~ G \, * \, {\Bar R} ~\to~ {\Bar U}$  \\ \hline
$ ~~G \, * \, R ~~$ &  $ ~~ G \, * \, {\Bar R} ~\to~ {\Bar R}
~~,~~ ~~ G \, * \, {\Bar U} ~\to~ {\Bar R}$  \\ \hline\hline
$ ~~T \, * \, L ~~$ &  $ ~~ T \, * \, {\Bar T} ~\to~ {\Bar L}
~~$  \\ \hline
$ ~~T \, * \, G ~~$ &  $ ~~ T \, * \, {\Bar G} ~\to~ {\Bar G}$
~~  \\ \hline
$ ~~T \, * \, T ~~$ &  $ ~~ T \, * \, {\Bar T} ~\to~ {\Bar T}
~~$  \\ \hline
$ ~~T \, * \, U ~~$ &  $ ~~ T \, * \, {\Bar U} ~\to~ {\Bar U}
~~$  \\ \hline
$ ~~T \, * \, R ~~$ &  $ ~~ T \, * \, {\Bar R} ~\to~ {\Bar R}
,~~ ~~ T \, * \, {\Bar U} ~\to~ {\Bar R}
$  \\ \hline \hline
$ ~~U \, * \, L ~~$ &  $ ~~ U \, * \, {\Bar U} ~\to~ {\Bar L}
~~$  \\ \hline
$ ~~U \, * \, G ~~$ &  $ ~~ U \, * \, {\Bar U} ~\to~ {\Bar G},  
~~ U \, * \, {\Bar R} ~\to~ {\Bar G}~~   $ \\ \hline
$ ~~U \, * \, T ~~$ &  $ ~~ U \, * \, {\Bar U} ~\to~ {\Bar T}
~~$  \\ \hline
$ ~~U \, * \, U ~~$ &  $ ~~ U \, * \, {\Bar U} ~\to~ {\Bar U}
 $  \\ \hline
\end{tabular}
\end{center}

\renewcommand\arraystretch{1.00}
\begin{center}
\begin{tabular}{|c|c|}\hline
\multicolumn{2}{|c|}{ \bf Table 2} \\ \hline\hline
${\rm {Commutator}}$ & ${\rm {Co-adjoint}}~{\rm {Action(s)}}$ 
\\ \hline 
$ ~~U \, * \, R ~~$ &  $~~ U \, * \, {\Bar R} ~\to~ {\Bar R},
~~ U \, * \, {\Bar U} ~\to~ {\Bar R}
$  \\ \hline
$ ~~R \, * \, L ~~$ &  $~~ R \, * \, {\Bar R} ~\to~ {\Bar L},
~~ R \, * \, {\Bar U} ~\to~ {\Bar L} ~~ 
~~$  \\ \hline
$ ~~R \, * \, G ~~$ &  $~~ R \, * \, {\Bar R} ~\to~ {\Bar G},
~~ R \, * \, {\Bar U} ~\to~ {\Bar G} ~~ 
~~ $ \\ \hline
$ ~~R \, * \, T ~~$ &  $ ~~ R \, * \, {\Bar R} ~\to~ {\Bar T},
~~ R \, * \, {\Bar U} ~\to~ {\Bar T}~~ 
~~$  \\ \hline
$ ~~R \, * \, U ~~$ & $~~ R \, * \, {\Bar R} ~\to~ {\Bar U},
~~ R \, * \, {\Bar U} ~\to~ {\Bar U} $  \\ \hline
$ ~~R \, * \, R ~~$ & $ ~~ R \, * \, {\Bar R} ~\to~ {\Bar R} $\\
\hline
\end{tabular}
\end{center}

From these we can see that ${\cal L}^{\prime} * {\Bar {\cal L}}$ will
lead to changes in the coadjoint vectors as:
$$ \eqalign{
\d {\Bar L} &=~ L \, *\, {\Bar L} ~+~ G \, *\, {\Bar G} ~+~  T \, 
*\,  {\Bar T} ~+~  U \, *\, {\Bar U} ~+~ R \, *\, {\Bar R} ~+~  R \, 
*\, {\Bar U} ~~~, \cr
\d {\Bar G} &=~ L \, *\, {\Bar G} ~+~ G \, *\, {\Bar L} ~+~  G \, 
*\,  {\Bar T} ~+~  G \, *\, {\Bar U} ~+~ T \, *\, {\Bar G} ~+~  U \, 
*\, {\Bar U} \cr 
&{~~~~}~+~ U \, *\, {\Bar R} ~+~  R \, *\, {\Bar R}
 ~+~ R \, *\, {\Bar U} ~~~, \cr 
\d {\Bar T} &=~ L \, *\, {\Bar T} ~+~ G \, *\, {\Bar G} ~+~  G \, 
*\,  {\Bar U} ~+~  T \, *\, {\Bar T} ~+~ U \, *\, {\Bar U} 
 ~+~ R \, *\, {\Bar R}\cr 
&{~~~~} ~+~  R \,  *\, {\Bar U} ~~~, \cr
\d {\Bar U} &=~ L \, *\, {\Bar U} ~+~ G \, *\, {\Bar U} ~+~  G \, 
*\,  {\Bar R} ~+~  T \, *\, {\Bar U} ~+~ U \, *\, {\Bar U} 
 ~+~ R \, *\, {\Bar R}\cr 
&{~~~~} ~+~  R \,  *\, {\Bar U} ~~~, \cr
\d {\Bar R} &=~ L \, *\, {\Bar R} ~+~ L \, *\, {\Bar U} ~+~  G \, 
*\,  {\Bar R} ~+~  G \, *\, {\Bar U} ~+~  T \,  *\,  {\Bar R}
 ~+~ T \, *\, {\Bar U} \cr 
&{~~~~} ~+~ U \, *\, {\Bar R} ~+~ U \, *\, {\Bar U} ~+~  R \,  
*\, {\Bar R} ~~~,\cr
\d{\bar \b} & = 0.
} \eqno(27)$$
\newline ${~~}$ \newline \noindent
{\bf {(III.c) Explicit Variations}}

{\center \bf{$L \, *  L$\,\, Commutator:}}

Starting with the invariant pairing we have:
$$ ( L_{a}, \a ) \, < \, ({\Bar L}{}_{\bar z}, {\bar \b}) 
\, | \, ({L}{}_{z}, \b) \, >
~=~ 0 ~~~. \eqno(28)$$
Then by Leibnitz rule,
$$
< \, ( L_{a}, \a ) \, * \, ({\Bar L}{}_{\bar z}, {\bar \b}) 
\, | \, ({L}{}_{z}, \b) \, > \,+\,  < \, ({\Bar L}{}_{\bar 
z}, {\bar \b}) \, | \, (L_{a}, \a ) \, *\, ({L}{}_{z}, \b) \, >
~=~ 0 ~~~. \eqno(29)
$$
Since we know the adjoint action we may write
$$
< \, ( L_{a}, \a ) \, * \, ({\Bar L}{}_{\bar z}, {\bar \b}) \, 
| \, ({L}{}_{z}, \b) \, > \,=\, -\, < \, ({\Bar L}{}_{\bar z}, 
{\bar \b}) \, | \, \Big( \, ( a \,- \, z) \, L_{a + z},  \,+\, 
\fracm 18 c (\, a^3 - a )\,  \d_{a + z, 0} \, \Big) \, >
\eqno(30)$$
which implies that,
$$
 < \, ( L_{a}, \a ) \, * \, ({\Bar L}{}_{\bar z}, {\bar \b}) 
\, | \, ({L}{}_{z}, \b) \, > ~=~ - \, \Big\{ ~  ( a \,- \, z) \, 
\d_{\bar z, a + z} ~+~ \fracm 18 c (\, a^3 - a  ) \, \bar \b \, 
\d_{a + z, 0} ~ \Big\}  ~~~,~~~
\, 
$$
$$
\to ~~  ( L_{a}, \a ) \, * \, ({\Bar L}{}_{\bar z}, {\bar \b}) 
\,=\,-\,  \Big(~  ( \, 2 a \,- \, \bar z\, ) \, {\Bar L}{}_{\bar 
z - a} \,+\, \fracm 18 \,c \, \bar \b \, (\, a^3 - a  ) \, {\Bar 
L}{}_{- a} ,\, 0  ~\Big)  ~~~,
\, \eqno(31)
$$
where $ z \,=\, \bar z - a $, so that
$$
( L_{a}, \a ) \, * \, ({\Bar L}{}_{\bar z}, {\bar \b}) \,=\, -\, 
\Big( \,  ( \, 2 a \,- \, \bar z\, ) \, {\Bar L}{}_{\bar z - a} 
\,+\, \fracm 18 \,c \, \bar \b \, (\, a^3 -  a  ) \, {\Bar L}{
}_{- a} ,\, 0  ~\Big) ~~~. \eqno(32)
$$
Rewriting in terms of functions instead of modes we have for 
functions $\xi$ and $D$,
$$
 ( L_{\xi}, \a ) \,*\, ({\Bar L}{}_{D}, {\bar \b})  ~=~ \, ({\Bar
     L}_{\tilde D},0) ~~~, \eqno(33)
$$
where ${\tilde D} =  -\,(\, 2 \xi^{\prime} D \, +\, \xi\, D^{\prime}
\,+\, \fracm 18 c \, \bar \b \,\xi^{\prime \prime \prime })$.  This 
shows the usual transformation of a quadratic differential, $D$, with 
respect to the vector field $\xi$.  Up to the inhomogeneous term 
$D$ transforms as a rank two tensor.  It is the inhomogeneous term 
that violates tensorality.  From the adjoint action one sees that 
$\xi$ transforms as a rank one contravariant tensor in one dimension 
making it easy to identify with $\xi^\a$ from a Lie derivative.  
In the same way $D$ can be thought of as a two index object $D_{\a
\b}$.   This suggests that a spin two type object is present in the 
spectrum.  Throughout we will treat the action of $\xi$ as a 
one dimensional Lie derivative (up to extensions) in order to 
understand the type of fields that are present in the dual.

{\center {\bf $G \, * L$ Commutator}:}

In the same way we examine the action of the $G_{b}{}^{\rm I} $ on 
the pairing.  Since the pairing is invariant we have 
$$
G_{b}{}^{\rm I}  \, < \,  {\Bar G}{}_{\bar y}{}^{{\Bar {\rm Q}}}  
\, | \, {L}{}_{z} \, > ~=~ 0 ~~~, \eqno(34)$$
by Leibnitz
$$
 < \, G_{b}{}^{\rm I}  \,* \,  {\Bar G}{}_{\bar y}{}^{{\Bar {\rm Q}}} 
\, | \, {L}{}_{z}
\, > ~+~  < \, {\Bar G}{}_{\bar y}{}^{{\Bar {\rm Q}}}  \, | \, 
G_{b}{}^{\rm I}  \,* \,
{L}{}_{z} \, > ~=~ 0 ~~~,\eqno(35)$$
which implies that 
$$ \eqalign{
 < \, G_{b}{}^{\rm I}  \,* \,  {\Bar G}{}_{\bar y}{}^{{\Bar {\rm Q}}} 
\, | \, {L}{}_{z} \, > &=~ - \,  < \, {\Bar G}{}_{\bar y}{}^{{\Bar 
{\rm Q}}}  \, | \, - \, (~ \fracm 12 z \, - \, b ~) \, G_{b}{}^{\rm I}   
\, >    ~=~
 (~ \fracm 12 z \, - \, b ~) \, \d^{{\rm I}\, {\bar {\rm Q}} } \,\,
\d_{\bar y, b + z} ~~,~~ 
}\eqno(36)$$
where $ z = \bar y - b $. This yields
$$
G_{b}{}^{\rm I}  \,* \,  {\Bar G}{}_{\bar y}{}^{{\Bar {\rm Q}}} ~=~
(\fracm 12 {\bar y} - \fracm 32 b) \, {\bar L}_{{\bar y}-b}\,\d^{{
\bar Q} I} ~~~~~. \eqno(37)$$
Rewriting in terms of functions we have 
we have that 
$$ G^I_{\chi^I} * {\bar G}^{\bar Q}_{\Psi^{\bar Q}} ~=~ {\bar L}_f ~~~,
\,\,\,\, {\rm where} \,\,\, f ~=~ (\, \fracm 12\,(\Psi^{\bar Q})' \chi^I 
- \fracm 32 (\chi^I)' \Psi^{\bar Q} \,) \, \d^{{\bar Q} I} ~~~.
\eqno(38) $$

{\center {\bf  All $*$ Commutators}:}
$$\eqalign{
L_\xi *( {\bar L}_{D},{\bar \b}) &=~ {\bar L}_{{\tilde D}}
 ~~~,~~~
{\tilde D} ~=~ -2 \,\xi' D \,-\, \xi \, D' \,-\, \fracm {c {\bar 
\beta}} 8 \, \xi''' ~~~,  \cr
L_\xi * {\bar G}^{\bar Q}_{\Psi^{\bar Q}} &=~  {\bar G}^{\bar Q}_{
\tilde {\Psi^{\bar Q}}} ~~~~,~~~ {\tilde {\Psi^{\bar Q}}} ~=~ 
-(\fracm 32 \xi' \psi^{\bar Q} + \xi (\psi^{\bar Q})') ~~~, \cr
L_\xi * {\bar T}^{\rm{\bar R} {\bar S}}_{\tau^{\rm{\bar R}{\bar S}}}  
&=~ {\bar T}^{\rm{\bar R} {\bar S}}_{{\tilde \tau}^{\rm{\bar R}{\bar 
S}}} ~~~,~~~ {\tilde \tau}^{\rm{\bar R}{\bar S}} \,=\, - \xi'
{\tau^{\rm{
\bar R}{\bar S}}} \,-\, \xi \, ({\tau^{\rm{\bar R}{\bar S}}})' ~~~, \cr
L_\xi * {\bar U}^{\rm {\bar V_1} \cdots {\bar V_n}}_{\o^{\rm {\bar
V_1} \cdots {\bar V_n}}} &=~ {\bar U}^{\rm {\bar V_1} \cdots {\bar 
V_n}}_{{\tilde \o}^{\rm {\bar V_1} \cdots {\bar V_n}}} \,+\, \fracm i2
(i)^{[\fracm{n-2}{2}]-[\fracm n2 ]}\, {\bar R}^{\rm [{\bar V_1} \cdots
{\bar V_{n-2}}}_{\xi''{\o^{\rm {\bar  V_1} \cdots {\bar V_n}}}}
\d^{\rm V_{n-1}] V_{n}} ~~~, \cr
&{~~~~~}~{{\tilde \o}^{\rm {\bar V_1} \cdots {\bar V_n}}} \,=\, 
(\fracm n2 -2) \, \xi' \, {\o^{\rm {\bar V_1} \cdots {\bar V_n}}} \,-\,  
\xi ({\o^{\rm {\bar V_1} \cdots {\bar V_n}}})' ~~~, \cr
L_\xi * {\bar R}^{\rm {\bar T}_1 \cdots {\bar T}_m}_{\rho^{\rm {\bar
T_1} \cdots {\bar T_m}}} &=~ {\bar R}^{\rm {\bar T_1} \cdots {\bar
T_m}}_{{\tilde \rho}^{\rm {\bar T_1} \cdots {\bar T_m}}} ~~~,~~~ \cr
&{~~~~~}~{{\tilde \rho}^{\rm {\bar T_1} \cdots {\bar T_m}}} = (\fracm 
{\rm m}2 -2) \, \xi' \, \rho^{\rm {\bar T_1} \cdots {\bar T_m}} - \xi 
\, (\rho^{\rm {\bar T_1} \cdots {\bar T_m}})' ~~~, \cr 
G^{\rm I}_{\chi^{\rm I}} * {\bar G}^{\rm \bar Q}_{\psi^{
\rm \bar Q}} &=~ \d^{\rm I {\bar Q}}\, {\bar L}_{\tilde \xi}+ 4{\bar 
T}^{\rm I {\bar Q}}_{(\chi^{\rm I} \psi^{\rm \bar Q})}~~~,~~~ {\tilde 
\xi}=\fracm 12 \, (\psi^{\bar Q})' \chi^I -\fracm 32 \, (\chi^I)' 
\psi^{\bar Q} ~~~, \cr
G^{\rm I}_{\chi^{\rm I}} * ({\bar L}_D, {\bar \b}) &=~ 4 i \, {\bar
G}^{\rm I}_{
(-\chi^{\rm I}\, D-{\bar \b} c (\chi^{\rm I})'')} ~~~,  \cr
G^{\rm I}_{\chi^{\rm I}} * {\bar T}^{\rm {\bar R\,S}}_{\tau^{\rm{\bar 
R\,S}}} &=~ \fracm i2 ({\bar G}{}^{\rm S}_{\chi^{\rm S}} \, \d^{\rm { 
\bar R} I} - {\bar G}{}^{\rm \bar R}_{\chi^{\rm \bar R}}\, \d^{\rm I 
{\bar S}}) ~~~, ~~~ \chi^{\rm{\bar R}} \,=\,\chi^{\rm S} = 2 (\chi^{\rm 
I})' \tau^{\rm{\bar R\,S}} + \chi^{\rm I} \, (\tau^{\rm {\bar R\,
  S}})'  
~~~, \cr
G^{\rm I}_{\chi^{\rm I}} * {\bar R}^{\rm {\bar T_1} \cdots {\bar
T_m}}_{\rho^{\rm {\bar T_1} \cdots {\bar T_m}}} &=~ 2 i (i)^{m+1}
(i)^{[\fracm{m+2}{2} ]-[\fracm m2 ]}\, {\bar U}^{\rm [{\bar T_1} 
\cdots {\bar T_m}]}_{(\chi^{\rm I} \r^{\rm {\bar T_1} \cdots {\bar
T_m}})}\cr
&\,\,\,\,\,\,\,\,\, - 2 i^{[{m-1\over 2}]-[\fracm {m-2}2]} \, 
\d^{I\,[{\bar T_1}}\,{\bar R}^{\rm {\bar T_2} \cdots {\bar T_m}]}_{
((\chi^{\rm I})' \r^{\rm {\bar T_1} \cdots {\bar T_m}}-(\chi^{\rm 
I}) (\r^{\rm {\bar T_1} \cdots {\bar T_m}})')}\cr
&\,\,\,\,\,\,\,\,\, - (i) (i)^{[\fracm {m+1}2 ]-[\fracm m2
]}\sum^{m+1}_{r=1}\,(-1)^{r-1}\, {\bar R}^{\rm  {\bar T_1} \cdots 
{\bar T_{r-1}}\, I \,{\bar T_{r+1}}\cdots {\bar T_m}}_{(\chi^{\rm 
I}\r^{\rm {\bar T_1} \cdots {\bar T_m}})} ~~~, \cr
T^{\rm J\, K}_{t^{\rm J\, K}} * {\bar G}^{\rm \bar Q}_{\psi^{\rm \bar 
Q}} &=~ -2 ({\bar G}^{\rm K}_{(t^{\rm J\, K}\, \psi^{\rm \bar Q})} 
\d^{\rm {\bar Q} J}-{\bar G}^{\rm J}_{(t^{\rm J\, K}\, \psi^{\rm \bar 
Q})} \d^{\rm {\bar Q} K}) ~~~,\cr 
G^{\rm I}_{\chi^{\rm I}} * {\bar U}^{\rm {\bar V_1}\cdots {\bar
V_n}}_{\o^{\rm {\bar V_1} \cdots {\bar V_n}}} &=~ - 2 i^{[{n-1\over
2}]-[\fracm n2]} \, \d^{I\,[{\bar V_1}}\,{\bar U}^{\rm {\bar V_2}
\cdots {\bar V_n}]}_{((n-4)(\chi^{\rm I})' \o^{\rm {\bar V_1} \cdots
{\bar V_n}}-(\chi^{\rm I}) (\o^{\rm {\bar V_1} \cdots {\bar
V_n}})')} \cr
&\,\,\,\,\,\,\,\,\,\, + 2(-1)^{n-1} (i)^{[\fracm {n-1}2]-[\fracm 
n2]} \,\d^{{\rm I}[ {\rm \bar V_n}} {\bar U}^{{\bar V_1}\cdots {\bar 
V_n}] \,{\rm K}}_{((\chi^{\rm I})' \o^{\rm {\bar V_1} \cdots {\bar
V_n}})}{}_{\rm K} \cr
&\,\,\,\,\,\,\,\,\,\, + (i) (i)^{[\fracm {n-1}2 ]-[\fracm n2
]}\sum^n_{r=1}\, {\bar U}^{\rm [ {\bar V_1} \cdots {\bar V_{r-1}}{
\bar V_{r+1}} \cdots {\bar V_n}}_{(\chi^{\rm I}\o^{\rm {\bar V_1} 
\cdots {\bar V_n}})}\,\d^{\rm{\bar V_r}\,]\,I}\cr
& \,\,\,\,\,\,\,\,\,\,+ {\bar G}^{ [ {\rm \bar V_2}}_{( - 4 i 
\, (\chi^{\rm I})' (\o^{\rm {\bar V_1}\cdots {\bar V_n}})'-2i\,
(\chi^{\rm I}) (\o^{\rm {\bar V_1}\cdots {\bar V_n}})'')} \,\d^{\rm 
\bar {V_3} \bar{V_4}}\, \d^{{\rm \bar V_1} ]{\rm I}}\, \d^{n 4} \cr
&\,\,\,\,\,\,\,\,\,\, -2i(-1)^n (i)^{[\fracm n2 ]-[\fracm{n-1}2
]}\, \d^{\rm I [ {\bar V_n}}\, {\bar R}_{(\chi^{\rm I} \o^{\rm {\bar 
V_1} \cdots {\bar V_n}})}^{\rm {\bar V_1}\cdots {\bar V_{n-1}}]} 
~~~, \,\cr
T^{\rm J\, K}_{t^{\rm J\, K}} * {\bar U}^{\rm {\bar V_1}\cdots {\bar
V_n}}_{\o^{\rm {\bar V_1} \cdots {\bar V_n}}} &=~ -\sum_{r=1}^{n-1} 
(-1)^{n+1} ( \d^{\rm J\, [ \, {\bar V_1}}\, {\bar U}^{\rm {\bar V_2}
\cdots {\bar V_{r-1}} | \,K\,| {\bar V_{r+1}} \cdots]{\bar V_n}}_{
(t^{\rm J\, K}\,\o^{\rm {\bar V_1} \cdots {\bar V_n}})} \cr 
&{~~~~~~~~~~~~~~~~~~~~~~~~} - \d^{\rm 
K\, [ \, {\bar V_1}}\, {\bar U}^{\rm {\bar V_2}\cdots {\bar V_{r-1}} 
| \,J\,| {\bar V_{r+1}} \cdots]{\bar V_n}}_{(t^{\rm J\, K}\,\o^{\rm 
{\bar V_1} \cdots {\bar V_n}})})\cr
&{~~~~~} + {\bar U}^{\rm [{\bar V_1}\cdots {\bar V_{n-1}}]\,J\,}_{
(t^{\rm J\, K}\,\o^{\rm {\bar V_1} \cdots {\bar V_n}})}\,\d^{\rm {\bar 
V_n}\,K} -{\bar U}^{\rm [{\bar V_1}\cdots {\bar V_{n-1}}]\,K\,}_{(t^{
\rm J\, K}\,\o^{\rm {\bar V_1} \cdots {\bar V_n}})}\,\d^{\rm {\bar
V_n}\,J}\cr
&{~~~~~}  -i (-1)^{n-2}\,(\d^{\rm K\, {\bar V_n}}\, \d^{J \,[ {\bar 
V_1}}\,-\, \d^{\rm J\, {\bar   V_n}}\,\d^{\rm K\,[{\bar V_1}}) {\bar 
R}^{\rm {\bar V_2}\cdots {\bar V_{n-1}}]}_{((t^{\rm J\, K})'\,\o^{\rm 
{\bar V_1} \cdots {\bar V_n}})} ~~~, \,  }$$
$$
\eqalign{ 
R^{\rm J_1\cdots J_p}_{r^{\{J_p\}}}* {\bar U}^{\rm {\bar V_1} 
\cdots {\bar V_m}}_{\o^{\{V_m\}}} &=~ - \fracm 12 i (i)^{\{[\fracm p2 
] - [\fracm {p+2}2]\}} \, \d_{\rm [{\bar J_1} \cdots {\bar J_p}]}^{\rm 
[ {\bar V_1} \cdots {\bar V_{m-2}} }\,\d^{\rm {\bar V}_{m-1}], {\bar 
V}_{m}}\, \d^{\rm m,p+2}\, {\bar L}_{(r \o)''}\cr
&{~~~~~} +  i^{\{[\fracm p2] -[\fracm{p+1}2]\}} \d^{\rm p+1,m} {\bar 
G}^{\rm \bar V_m}_{(r \o)'}\, \d^{\rm [ {\bar V_1}\cdots {\bar V_{m-1
}}]}_{\rm [ J_1 \cdots J_p]}  \cr
&{~~~~~} + 2 i (-1)^{p+1} \d^{\rm p+3,m} \,(i)^{[\fracm p2] - [\fracm 
{p+2}2]} {\bar G}^{[ {\bar V_1}}_{(r \o)''} \, \d^{\rm {\bar V_2} \cdots 
{\bar V_{m-2}}}_{\rm [ J_1 \cdots J_p]} \,\d^{\rm {\bar V_{m-1},\, ]
{\bar V_m}} } \cr
&{~~~~~} + i (-1)^p {\bar T}^{\rm {\bar V_1}[ {\bar V_m}}_{(r \o)'} \d^{
\rm {\bar V_2}\cdots {\bar V_{m-1}}]}_{\rm [ J_1 \cdots J_p ]} \d^{p+2,m
}\cr
&{~~~~~} + (-1)^{\rm pm} (i)^{\{[\fracm m2]+[\fracm p2]-[\fracm{m+p-2}2]
\}} \, {\bar U}^{\rm {\bar V_p+1} \cdots {\bar V_{m-p}}}_{(r \o)'} 
\d^{\rm {\bar V_1}\cdots {\bar V_{p}}}_{\rm [J_1 \cdots J_p]} ~~~, \cr
U^{\rm I_1 \cdots I_q}_{\mu^{\rm \{I_q\}}}*{\bar U}^{\rm {\bar V_1} 
\cdots {\bar V_m}}_{\o^{\rm \{V_m\}}} &=~ -\d^{\rm m\,q}\,\d^{\rm 
[\,I_1 \cdots I_q\,]}_{\rm [{\bar V_1}\cdots {\bar V_q}\,]} {\bar 
L}_{((\fracm {4-q}2) \mu \o-(\fracm {q-2}2) \mu \o')}\, - 2\, {\bar 
T}^{\rm {\bar V_m} J_q}_{(\o \mu)} \d^{\rm [\, I_1 \cdots I_q\,]}_{\rm 
[{\bar V_1}\cdots {\bar V_q}\,]}\,\d^{\rm q}_{\rm m} \cr
&{~~~~~} - 2i^{([\fracm {\bar V_1}2]-[\fracm {{\bar V_1}+1}2])}\,
\d^{\rm 
m, (q+1)} {\bar G}^{\rm [ {\bar V_1}}_{(-(q-2)\o' \mu+(3-q)\o \mu') 
}\, \d^{\rm {\bar V_2} \cdots {\bar V_m}\,]}_{\rm [ I_1 \cdots I_q ]}
\cr
&{~~~~~} + 2 (-1)^q (i)^{[\fracm q2 ]-[\fracm {q+1}2]} {\bar G}^{\rm
  I_q}_{(-\o \mu)'}\,\d^{\rm m,q+1} \, \delta^{\rm [ {\bar V_1}\cdots 
{\bar V_{m-2}}}_{\rm [I_1 \cdots I_{q-1}]}\, \d^{\rm {\bar V_{m-1}}], 
{\bar V_m}}\cr
&{~~~~~}- i (i)^{[\fracm q2 ]-[ \fracm {q-1}2]}\sum_{r=1}^{q-1} (-1)^{
r-1} \d^{\rm m}_{\rm q-1}\, {\bar G}^{\rm [ I_r}_{(\o \mu)} \d_{\rm \bar 
[ V_1}^{\rm I_1}\cdots \d_{\rm \bar V_{r-1}}^{\rm I_{r-1}} \, \d_{\rm 
\bar V_r}^{\rm I_{r+1}}\cdots \d_{\rm \bar V_m]}^{\rm I_q ]}
\cr
&{~~~~~} + \sum_{r=1}^{q-1} 2 (-1)^{r+1} ( {\bar T}^{\rm J_r [{\bar V_1}
}_{(\o \mu)} \d_{\rm [ I_{1}}^{{\bar V}_2} \cdots  \d_{{\rm I_{r-1}
}}^{{\bar V}_r} \d_{{\rm I_{r+1}}}^{{\bar V}_{r+1}} \cdots \d_{{\rm 
] I_{q}}}^{{] \bar V}_m} \d^{\rm q,m}) \cr
&{~~~~~}+ i (i)^{\{[\fracm q2]+[\fracm {m-q}2 +2]-[\fracm {m+2}2]\}}
\cr &{~~~~~~~~}
\times \{ \sum_{r=1}^{q}(-1)^{r-1}\d_{\rm  [\,I_1 \cdots I_{q-1}\,]
}^{\rm  [{\bar V_1}\cdots {\bar V_{q-1}}\,}\ {\bar U}^{\rm {\bar V_q} 
\cdots {\bar V_{q+r-1}} I_q {\bar V_{q+r}}\cdots {\bar ] V_m}}_{\o \mu} 
\cr
&{~~~~~} -(-1)^{q(m-q+2)}\sum_{r=1}^q (-1)^{r-1} {\bar U}^{\rm {\bar
V_1}
\cdots {\bar V_{m-q+1}} [\, I_r}_{\o \mu}  \d^{\rm I_1 \cdots I_{r-1}
I_{r+1}\cdots \, I_q ]}_{\rm {\bar V_{m-q+2}}\cdots{\bar V_{m-q+2+r}}
\cdots{\bar V_m}}~\}\cr
&{~~~~~} -(i)^{\{[\fracm q2] +[\fracm {m-q}2] -[\fracm {q+m-4}2]\}}
\, {\bar R}^{\rm [{\bar V_1} \cdots {\bar V_{m-q}}}_{(\o \mu)'} \, 
\d_{\rm [\,I_1 \cdots ] I_{q}\,}^{\rm {\bar V_{m-q+1}\cdots ] {\bar
V_{m}}\,}} ~~~, \cr
R^{\rm J_1\cdots J_p}_{r^{\{J_p\}}} * {\bar R}^{\rm {\bar T_1}\cdots
{\bar T_m}}_{\r^{\{T_m\}}} &=~ \d^{\rm [{\bar T_1} \cdots {\bar T_m
}]}_{\rm [ J_1 \cdots J_p ]}\, \d^{\rm p\,m} \, {\bar L}{}_{(-(\fracm 
p2 -2) r' \r - (\fracm p2 - 1)\,r \r')} \cr
&{~~~~} +(-1)^p\, \{2(i) {\bar G}^{\rm [ {\bar T_1}}_{((2-p) r' \r
-(p-1) r
  \r')} \, \d^{\rm {\bar T_2} \cdots {\bar T_m}]}_{\rm [J_1\cdots
  J_p]} \d^m_{p+1} 
\cr
&{~~~~} + (i)(i)^{[\fracm p2]-[\fracm {p-1}2]} \sum_{r=1}^{p} (-1)^r
{\bar
G}^{\rm J_r}_{(r \r)} \d^{\rm m}_{\rm p-1} \d^{[\rm J_1\cdots J_{r-1}
J_{r+1}\cdots J_p]}_{\rm [ {\bar T_1} \cdots {\bar T_{r-1}} {\bar
T_{r}}\dots {\bar T_m}]}\} \cr
&{~~~~} + \sum_{r=1}^{p} (-1)^{r+1} 2 {\bar T}^{\rm [ {\bar T_1}\,|J_r|
}_{(r \r)} \d^{\rm {\bar T_2} \cdots {\bar T_m} ]}_{\rm [ J_1\cdots
J_{r-1} \, J_{r+1} \cdots J_p]} \d^p_m \cr
&{~~~~} - \sum_{r=1}^p (-1)^{r-1} {\bar U}^{\rm [ {\bar T_1} \cdots {
\bar T_{m-p+1}} \, |J_r|}_{(r \r)} \d^{\rm {\bar T}_{m-p+2}\cdots {\bar
T}_m ]}_{\rm J_1 \cdots J_{r-1} J_{r+1} \cdots J_{p}} \cr
&{~~~~} + i^{\{[\fracm p2]+[\fracm{m-p}2]-[\fracm m2]\}} {\bar 
R}^{\rm [ {\bar T_{p+1}} \cdots {\bar T_m} }_{(2r' \r + r \r')}
\d^{\rm {\bar T_1} \cdots {\bar T_p} ]}_{\rm [ J_1 \cdots J_p]} ~~~, \cr
}$$
$$\eqalign{ 
T^{\rm J\, K}_{t^{\rm J\, K}} * {\bar R}^{\rm {\bar T_1}\cdots {\bar
T_m}}_{\r^{\rm {\bar T_1}\cdots {\bar T_m}}} &=~ \sum_{r=1}^{m}
\,(-1)^{r+1}\,(\d^{\rm [ {\bar T_1} |\,J\,|} \,{\bar R}^{\rm {\bar
T_2}\cdots {\bar T}_{r-1} \,|\,K\,|\,{\bar T_{r+1}}\cdots {\bar
T_m}]}_{(t^{\rm J\, K}\,\r^{\rm {\bar V_1} \cdots {\bar V_n}})}
\cr 
&{~~~~~~~~~~~~~~~~~~~~~~~~}  -\d^{\rm [ {\bar T_1} |\,K\,|} \,{\bar 
R}^{\rm {\bar T_2} \cdots {\bar T}_{r-1} \,|\,J\,|\,{\bar T_{r+1}}
\cdots {\bar T_m}]}_{(t^{\rm J\, K}\,\r^{\rm {\bar V_1} \cdots 
{\bar V_n}})}) ~~~, \cr
U^{\rm I_1 \cdots I_q}_{\mu^{\rm \{I_q\}}} *{\bar R}^{\rm {\bar 
T_1} \cdots {\bar T_m}}_{\r^{\rm \{T_m\}}} &=~ -i (-1)^{q(m-q+2)}
(i)^{\{[\fracm{m-q}2+2]+[\fracm q2]-[\fracm m2]\}}\,\cr
&{~~~~~~} \times \sum_{r=1}^{m-q+2}\, (-1)^{r-1}\d^{\rm [\,I_1 
\cdots I_{q-1}\,] }_{\rm [{\bar T_{1}\cdots {\bar T_{q-1}}\,]}} 
{\bar R}^{\rm {\bar T_1} \cdots {\bar T_{q+r-1}\, I_q \,{\bar 
T_{q+r+1}\cdots {\bar T_{m}}}}}_{\r \mu} ~~~, \cr
T^{\rm J\, K}_{t^{\rm J\, K}} * ({\bar T}^{\rm {\bar R}{\bar S}}_{
\tau^{\rm {\bar R}{\bar S}}}, {\bar \b}) &=~ \fracm 12 (\d^{\rm {\bar R}
J}
\d^{\rm {\bar S} K} -  \d^{\rm {\bar R} K} \d^{\rm {\bar S} J})\, {
\bar L}_{((t^{\rm J K})' \, \tau^{\rm {\bar R}\,{\bar S}})} +  \fracm 
12 {\bar T}^{\rm A\,B}_{(t^{\rm J\, K}\, \tau^{\rm {\bar R}{\bar S}})} 
\, \d^{\rm J K {\bar R} {\bar S}}_{\rm A B} + 4 {\bar \b} {\bar T}^{\rm
J K}_{
(\t^{\rm {\bar R}{\bar S}})'} ~~~, \cr 
&{~~~~~~}{\rm where} \,\,\,\,\,\,\d^{\rm J K {\bar R} {\bar S}}_{\rm A 
B} \equiv (\,{ \d^{\rm A K} \d^{\rm B {\bar S}} \d^{\rm {\bar R} J} -
\d^{\rm A K} 
\d^{\rm B {\bar R}} \d^{\rm {\bar S} J} + \d^{\rm A {\bar S}} \d^{\rm B
J}
\d^{\rm {\bar R} K} }   \cr
&{~~~~~~}{~~~~~~}{~~~~~~}{~~~~~~}{~~~~~} - \d^{\rm A {\bar R}} \d^{\rm
  J S} 
\d^{\rm {\bar S} K} + {\rm \d^{A {\bar S}} \d^{\rm B K} \d^{\rm {\bar R}
J} - 
\d^{\rm A {\bar R}} \d^{\rm K B} \d^{\rm {\bar S} J} } \cr
&{~~~~~~}{~~~~~~}{~~~~~~}{~~~~~~}{~~~~~} + \d^{\rm A J} \d^{\rm B {\bar
S}} 
\d^{\rm { \bar R} K} - \d^{\rm J  A} \d^{\rm {\bar R} B} \d^{\rm {\bar
S} K}
+ {\rm \d^{A {\bar S}} \d^{B K} \d^{{\bar R} J}  } \, )~~~,
  }\eqno(39) $$
where the symmetry of the indices on the left hand side should be
imposed on the indices on the right side.  In the above, we have 
sometimes suppressed the indices associated with the functions used 
by the generators.   For example $ {\o^{\rm {\bar V_1} \cdots {\bar 
V_n}}}$ the associate of the ${\bar U}$ dual element may be written 
as $\o^{\{\rm V_m\}}$ or simply as $\o$. Also the notation $\d^{\rm 
I_1 \cdots I_m}_{\rm J_1 \cdots J_m} \equiv \d^{\rm I_1}_{\rm J_1} 
\cdots \d^{\rm I_m}_{\rm J_m}$ was utilized.  

There is quite a bit of interchange between functions in various sectors 
suggesting that the inhomogeneous contribution in the transformation
laws for $D$, is the interaction of the central extension with $D$.   
The coadjoint of the Virasoro algebra, i.e. the action of L on the
coadjoint  vectors reveals a spectrum of states containing: 
\begin{itemize}
\item $D$ corresponds to a rank 2 covariant tensor when the central
extension is set to zero and is otherwise a quadratic differential.
\item $\psi^{\rm \bar I}$ corresponds to $N$ spin-$\fracm 32$ fields
that partner with $D$. 
\item $\tau^{\rm {\bar R}{\bar S}}$ corresponds to the spin-1
covariant tensors that serves as the $N (N - 1)/2$ SO($N$) gauge 
potentials associated with the supersymmetries. 
\item Given the $N$ supersymmetries there are the fields 
 $ {\o^{\rm {\bar V_1} \cdots {\bar V_p}}}$. For a fixed
 value of $N$, the total number of independent components is
 given by
$$
\#(U) ~=~  N \, (\, 2^N \,-\, N -1\,) ~~~.
$$
\item Again given $N$ supersymmetries, there are the fields 
$ {\r^{\rm {\bar T_1} \cdots {\bar T_p}}}$.  For a fixed value of 
$N$, the total number of independent components is given by
$$
\#(R) ~=~  (\, 2^N \,-\, N -1 \,) ~~~.
$$
\end{itemize}
The spins of the fields associated with $U$ and $R$ vary according 
to $(2-\fracm p2)$.  These likely correspond to other gauge and
non-gauge physical fields, auxiliary, and Stuckelberg fields that 
are required to close the supersymmetry algebra.

We end our discussion with a conjecture.  If M-theory possesses a 
1D NSR formulation,  it seems likely that the $N$ = 32 or 16 case of the
present discussion determines the structure of its representation. We
conjecture that the spectrum of the 1D, $N$ = 32 or 16 ${\cal GR}$ super
Virasoro theory provides a set of  fundamental NSR variables to describe
M-theory.

${~~~~~~~~~~~~~~~~~~~~~~~~~~~~~~~~~~~~~~~~~~~~~~~~}$
${~~~~~~~~~~~~~~~~~~~~~~~~~~~~~~~~}$ 

${~~~~~~~~~}$ ``{\it {Equations were drawn up in paisley form.}}'' --
Rakim (1997)
$${~~~}$$
\noindent
{\Large {\bf {Acknowledgment}}} \newline \noindent
${~~~~~~~~~~~~~~~~~~~~~~~~~~~~~~~~~~~~~~~~~~~~~~~~}$
${~~~~~~~~~~~~~~~~~~~~~~~~~~~~~~~~~~~~~~~~~~~~~~~~}$
${~~~~~~~~~~~~~~~~~~~~~~~~~~~~~~~~~~~~~~~~~~~~~~~~}$
${~~~~}$ We thank Takeshi Yasuda for discussion.  V.G.J.R. and
C.C. thank the University of Maryland for hospitality where
this work was initiated.  S.J.G. likewise acknowledges the 
University of Iowa for hospitality during the completion of this 
work.  This work was supported in part by NSF grant \# PHY-96-43219.

\end{document}